\documentclass{article}

\usepackage{arxiv}

\usepackage[utf8]{inputenc} 
\usepackage[T1]{fontenc}    
\usepackage{hyperref}       
\usepackage{url}            
\usepackage{booktabs}       
\usepackage{amsfonts}       
\usepackage{nicefrac}       
\usepackage{microtype}      
\usepackage{lipsum}
\usepackage{graphicx}
\graphicspath{ {./images/} }
\usepackage{multirow} 
\usepackage{float}
 \usepackage{threeparttable}
 \usepackage{array}
\usepackage{algorithm}
\usepackage{algorithmic}
\usepackage{listings}
\usepackage{xcolor}
\usepackage{listings}
\usepackage{xcolor}
\usepackage{amsmath} 
\usepackage{subfigure}

\title{AutoFeedback: An LLM-based Framework for Efficient and Accurate API Request Generation}

\author{
 Huanxi Liu \\
  College of Computing\\
   National University of Defense Technology\\
  Changsha, China \\
  \texttt{liuhuanxi18@nudt.edu.cn} \\
\And
 Jiaqi Liao \\
  College of Computing\\
   National University of Defense Technology\\
  Changsha, China \\
  \texttt{liaojiaqi2019@163.com} \\
\And
 Dawei Feng \\
  College of Computing\\
   National University of Defense Technology\\
  Changsha, China \\
  \texttt{davyfeng.c@qq.com} \\
\And
 Kele Xu \\
  College of Computing\\
   National University of Defense Technology\\
  Changsha, China \\
  \texttt{kelele.xu@gmail.com} \\
\And
 Huaimin Wang \\
  College of Computing\\
   National University of Defense Technology\\
  Changsha, China \\
  \texttt{whm\_w@163.com} \\
}

\begin{document}
\maketitle
\begin{abstract}
Large Language Models (LLMs) leverage external tools primarily through generating the API request to enhance task completion efficiency. The accuracy of API request generation significantly determines the capability of LLMs to accomplish tasks.
  Due to the inherent hallucinations within the LLM, it is difficult to efficiently and accurately generate the correct API request.
  Current research uses prompt-based feedback to facilitate the LLM-based API request generation. However, existing methods lack factual information and are insufficiently detailed.
  To address these issues, we propose AutoFeedback, an LLM-based framework for efficient and accurate API request generation, with a Static Scanning Component (SSC) and a Dynamic Analysis Component (DAC). SSC incorporates errors detected in the API requests as pseudo-facts into the feedback, enriching the factual information. DAC retrieves information from API documentation, enhancing the level of detail in feedback. 
  Based on this two components, Autofeedback implementes two feedback loops during the process of generating API requests by the LLM.
   Extensive experiments demonstrate that it significantly improves accuracy of API request generation and reduces the interaction cost. AutoFeedback achieves an accuracy of 100.00\% on a real-world API dataset and reduces the cost of interaction with GPT-3.5 Turbo by 23.44\%, and GPT-4 Turbo by 11.85\%.
\end{abstract}


\section{Introduction}
Large Language Models (LLMs) are becoming increasingly intelligent and autonomous, progressing towards solving real-world pragmatic tasks \cite{saycan}\cite{gorilla}\cite{restgpt}\cite{toolllm}\cite{hugginggpt}.
By combining the impressive intention comprehension ability with \textit{external tools}, such as Application Programming Interfaces (APIs) \cite{restgpt}\cite{toolllm}\cite{toolalpaca} and code interpreters\cite{toolqa}\cite{agentbench}, LLMs have further extended their application scenarios. 

Notwithstanding substantial advancements in tool-augmented LLMs applied in research assistant \cite{boiko2023autonomous}\cite{m2024augmenting}, and software development \cite{metagpt}\cite{swebench}. The potentiality inherent in the integration of LLMs with external tools remain predominantly underexplored. Even one of the existing state-of-the-art models (GPT-4 \cite{gpt4}) still struggle to use tools to solve real-world problems\cite{restgpt}\cite{toolllm}\cite{apibank}. One of the primary problems is that the LLM is susceptible to hallucinations \cite{hallucination} - generating plausible yet ungrounded information. This poses great challenges in using external tool that usually requires standardized format input and accurate parameter values.

\begin{figure}[htbp]
  \centering
  \includegraphics[width=0.5\textwidth]{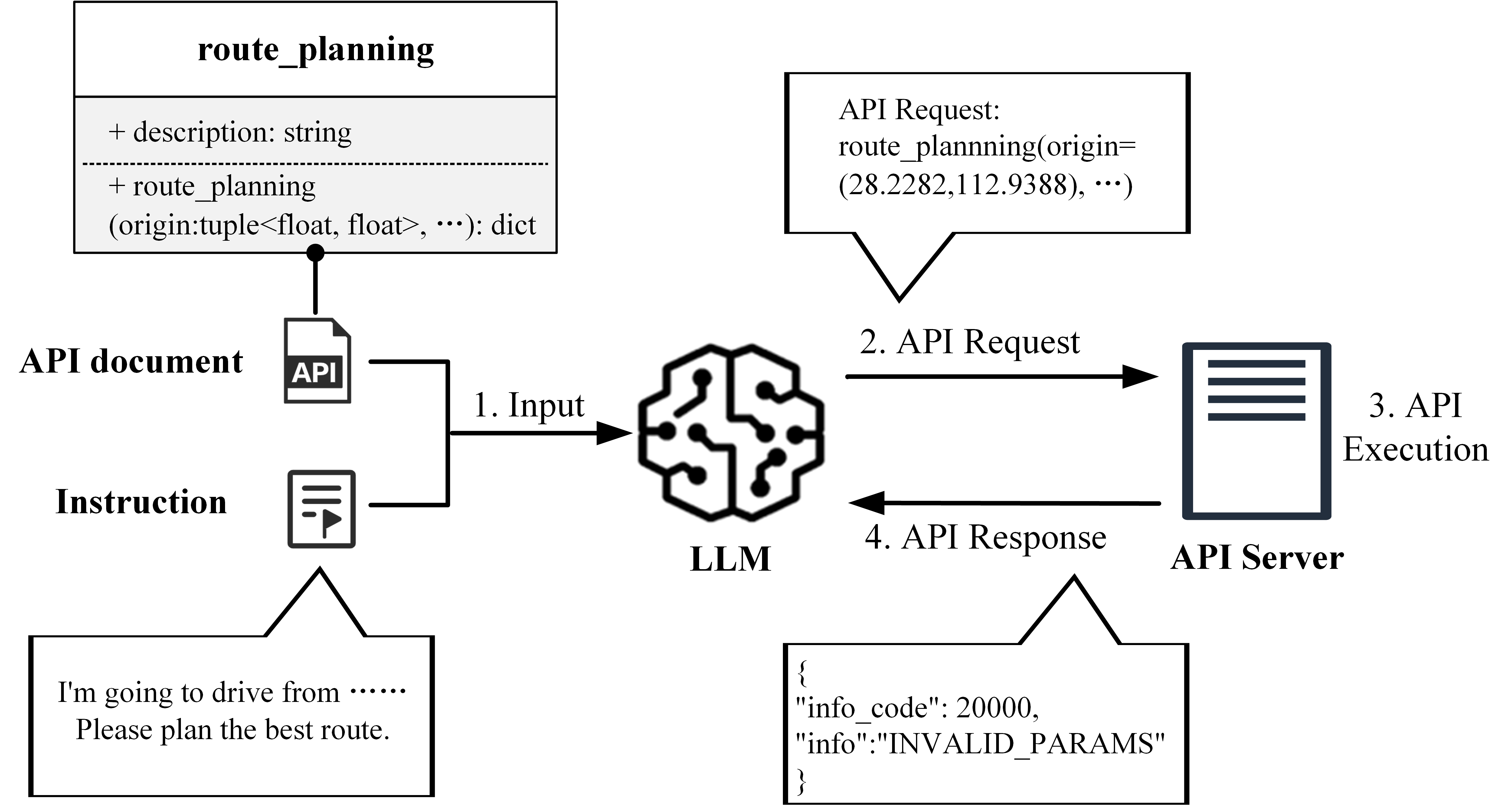}
  \caption{\small An example of LLM using external tool.}
  \label{figure-example}
  \vspace{-0.2cm}
\end{figure}


To reduce the hallucinations, fine-tuning \cite{gorilla}\cite{toolalpaca}\cite{toolformer} on particular datasets has been proven to be an effective approach. While this approach typically works for specific tools within the dataset, it is hard to generalize, not to mention the high labor costs associated with dataset construction.

Other existing research provides linguistic prompts to stimulate the planning and reasoning ability of LLMs \cite{riseagent}.
For example, task demonstrations\cite{shot}, self-thoughts\cite{wei2022chain}, self-examination \cite{selfcheck}\cite{deliberate} are used as static feedback (no direct interaction with external environment), to complete user's task.
However, instead of providing factual information such as API call format error and runtime execution error, static feedback predominantly comes from the output of LLMs, potentially exacerbating the phenomenon of hallucinations. Further, methods interacting with external environment provide dynamic feedback to facilitate the task completion process \cite{react}\cite{rag}. However, such feedback lacks sufficient information, resulting in less accuracy in guiding the LLM.

\begin{table}[t]
  \caption{Statistics and Parameter Number of External Tools Datasets.}
    \label{tab:dataset}

  \resizebox{\linewidth}{!}{
\begin{tabular}{c |cccc|ccccc}
\toprule\toprule
\multirow{2}{*}{\textbf{\large{Dataset}}} & \multicolumn{4}{c|}{\textbf{Statistics}}                                    & \multicolumn{5}{c}{\textbf{Parameter Number}}                                                            \\ 
                         & \footnotesize{\# of domains} & \footnotesize{\# of APIs} & \footnotesize{of samples}  & \footnotesize{avg. steps}  & \footnotesize{\# of string} & \footnotesize{\# of int \& float} & \footnotesize{\# of list \& tuple} & \footnotesize{\# of dict \& bool} & \footnotesize{avg. num.} \\ \midrule
\small{API-Bank \cite{apibank}}                 & 1000          & 2138       & 3146       & 1.00            & 5401         & 974                & 132                 & 53                  & 2.18        \\
\small{MP-API\footnotemark[2]}                   & 1             & 26         & 379             & 1.00        & 131          & 508                & 802                 & 0                & 3.80        \\
\footnotesize{ToolAlpaca-single \cite{toolalpaca}}        & 11            & 40         & 116              & 1.00        & 218          & 11                 & 2                   & 0                 & 1.99        \\
\footnotesize{ToolAlpaca-mix \cite{toolalpaca}}           & 20            & 94         & 134              & 1.38       & 509          & 85                 & 11                  & 0                   & 4.51        \\ \midrule
\end{tabular}
}
\vspace{-0.3cm}
\end{table}

We use a real-world example to illustrate the deficiency in dynamic feedback, as shown in Figure \ref{figure-example}. The LLM assists users in planning their driving routes through the \textit{route\_planning} API, but it incorrectly reverses the order of latitude and longitude, causing failure API execution. And the API response only contain an "info\_code: 20000" as a brief message, which is imprecise for LLMs to fix API error.


To alleviate these issue, we first investigate the API request generation of different LLMs on four datasets, and summarize the following four main error types as the factual information in the feedback. We provide motivating study for each error in Section 2:
\begin{itemize}
    \item[1)] \textbf{E1: NO\_API: } 
    API requests are not generated or the generated is failed to be parsed due to incorrect format.  
     \item[2)] \textbf{E2: API\_NAME\_MISMATCH: }
It could be a formatting mismatch between the generated API name with the API documentation, or a mismatch with the user instruction.
    \item[3)] \textbf{E3: PARAMETER\_INVALID: } 
    The generated output contains invalid parameter names which doesn't appear in the API documentation.
    \item[4)] \textbf{E4: INPUT\_MISMATCH: } 
    The types of parameter values are mismatched with the API documentation, or a mismatch with the user instructions.
\end{itemize}

To mitigate errors in generating API requests for LLMs, we propose AutoFeedback\footnote{\url{https://anonymous.4open.science/r/AutoFeedback-BCD0}}, an LLM-based framework, automatically provides feedback and corrects the four main errors. Specifically, AutoFeedback consists of two main components: (1) Static Scanning Component (SSC) and (2) Dynamic Analysis Component (DAC). SSC strengthens the reliability of static feedback by locally scanning API requests and categorizing detected errors into our predefined categories; DAC enhances the effectiveness of dynamic feedback by providing the API response with its comprehensive error message. For issues that AutoFeedback cannot solve, users could refer to the logs generated during the feedback process to manually provide more specific feedback or other methods.

We conduct an extensive evaluation of AutoFeedback on four datasets. Results demonstrate that AutoFeedback significantly improves the accuracy of LLM-based API request generation, achieving an accuracy of 100.00\% on a real-world API dataset, outperforms existing methods; AutoFeedback also substantially reduces the cost of accurate interaction, showing a decrease of 23.44\% on GPT-3.5 Turbo \cite{gpt3.5}, and 11.85\% on GPT-4 Turbo \cite{gpt4}.


To summarize, the contributions of this paper are:
\begin{itemize}
    \item Through empirical analysis, we propose four main error types in LLM-based API request generation, and provide them as factual information feedback to the LLM.

    \item We present AutoFeedback, a widely-applicable automated LLM-based framework, with two components for feedback LLMs to efficiently and accurately generate API requests. 

    \item Experiments demonstrate that AutoFeedback significantly improves the accuracy of LLM-based API request generation for task completion and reduces the cost of accurate interaction. 
\end{itemize}

\begin{figure}[htbp]
\setlength{\abovecaptionskip}{0cm}
\setlength{\belowcaptionskip}{-0.cm}
  \centering
  \includegraphics[width=0.5\linewidth]{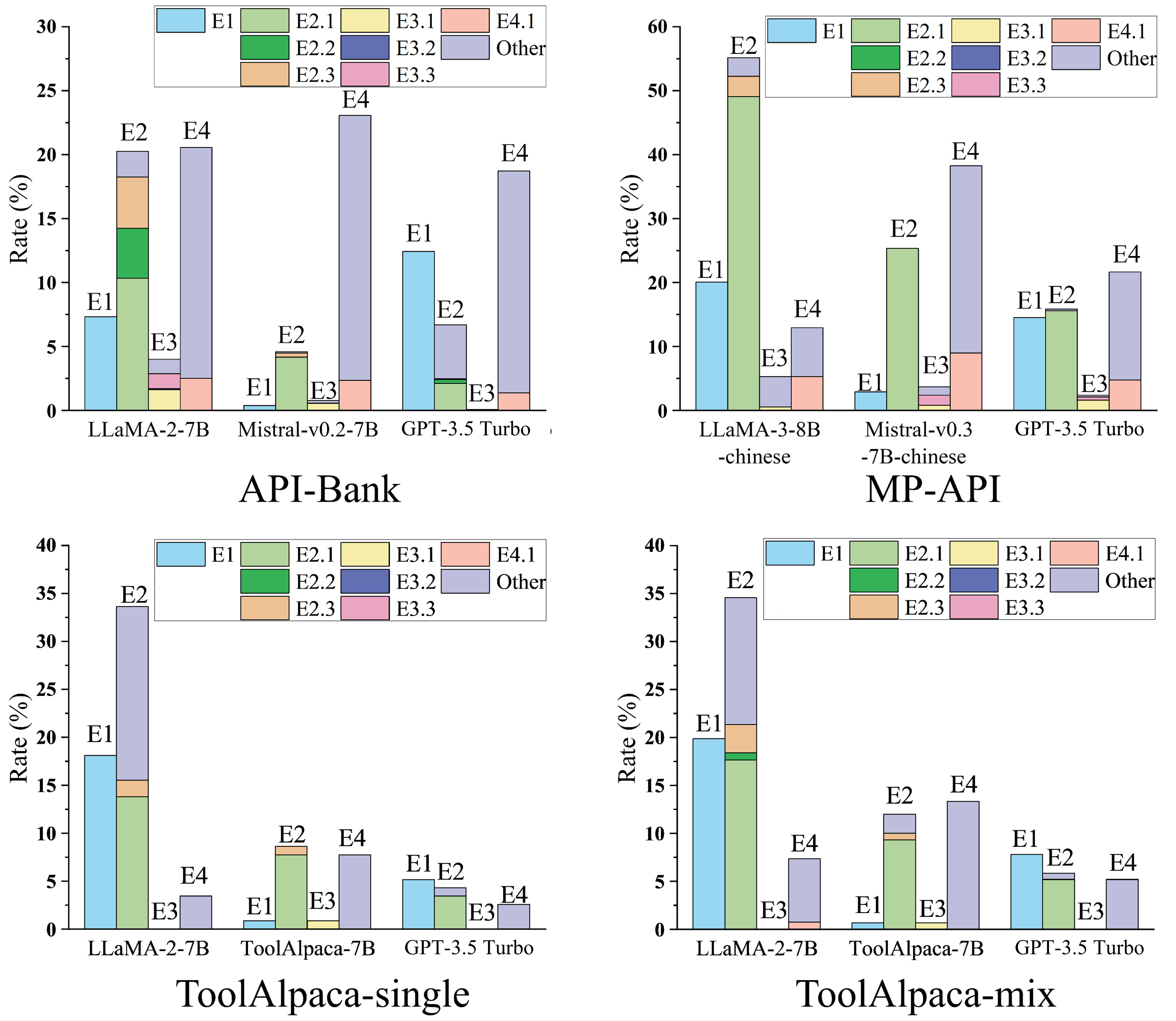}
  \caption{\small{Error Distribution of API requests on datasets.}}
  \label{error case}
\vspace{-0.3cm}
\end{figure}

\section{MOTIVATING SYUDY}
In this section, we provide detailed motivating examples to illustrate the four errors mentioned in Section 1.
\subsection{API Request Generation}
Taking user's instructions as input, LLMs generate API requests based on the API documentation. However, due to the hallucinations, the LLM has difficulty in generating correct API requests and cannot meet user requirements. We conducted research on various errors of the generated API request across four datasets. API-Bank \cite{apibank} is a famous tool-using synthetic dataset with 2138 APIs, 3146 samples and 1000 domains which determines the functionality of the APIs, such as healthcare and fitness; Materials Project API (MP-API) is a code dataset on Chinese instructions with 26 APIs and 379 samples; ToolAlpaca-single and -mix datasets \cite{toolalpaca} are open-source datasets of real-world API tools. The general information about datasets is shown in the Table \ref{tab:dataset}, and 
We will provide other detailed descriptions about datasets in the experiment section.

To exemplify the robustness, our practice covers models of different architectures with open and closed source, fine-tuned and non-fine-tuned.
We used the following LLMs for the study: GPT-3.5 Turbo \cite{gpt3.5}, LLaMA-2-7B \cite{llama2}, Mistral-V0.2-7B \cite{mistral}, and ToolAlpaca-7B\footnote{\url{https://huggingface.co/TangQiaoYu/ToolAlpaca-7B}} \cite{toolalpaca}. Besides, for Chinese dataset MP-API, We used the latest open-source two models: LLaMA-3-8B-chinese\footnote{{\url{https://github.com/ymcui/Chinese-LLaMA-Alpaca-3}}}, Mistral-V0.3-7B-chinese\footnote{\url{https://huggingface.co/shenzhi-wang/Mistral-7B-v0.3-Chinese-Chat}}. A detailed description of these LLMs will be also presented in the experiment section. 

\begin{figure}[htbp]
\setlength{\abovecaptionskip}{0cm}
\setlength{\belowcaptionskip}{-0.cm}
  \centering
  \includegraphics[width=0.95\linewidth]{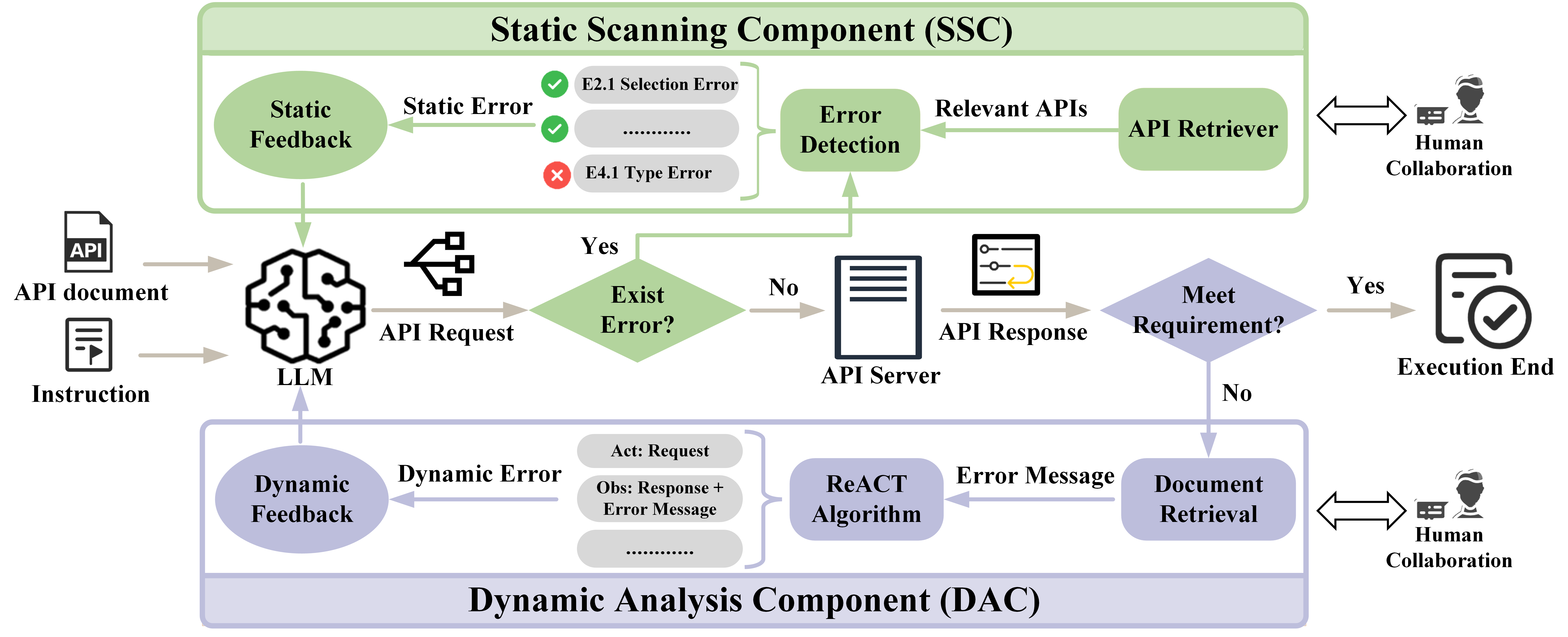}
  \caption{\small{The overview framework of AutoFeedback with two components: Static Scanning Component and Dynamic Analysis Component.}}
  \label{fig:framework}
\vspace{-0.3cm}
\end{figure}

\subsection{Error Analysis}
The results of the API error cases are shown in Figure \ref{error case}. 
We categorize errors based on their location in API requests, as the structural information contained within the error categories is factual, which could be subsequently fed back to the LLM.
The generated API request is compared with the ground true to determine its error type, and the order of our judgement process is \textbf{E1->E2->E3->E4}. The results indicate that all of LLMs are prone to produces incorrect output in the API request generation.

To increase factual information in the error categories, we further decompose these major errors into sub-errors.
If the LLM is unable to generate the correct API format, it is classified as \textbf{E1}. With regards to E2, we found the sources fall into the following three main dimensions: \textbf{E2.1} Selection Error: calling other candidate API name; \textbf{E2.2} Literal Error: incorrect letter case, or different naming methods (camel method, underscore method) in API name; \textbf{E2.3} Semantic Error: semantically similar to the correct API name. 
Similar to E2, the sources of E3 also come from these three dimensions: \textbf{E3.1}. Selection Error: calling parameter name in other candidate API; \textbf{E2.2} Literal Error: incorrect letter case, or different naming methods in parameter name; \textbf{E2.3} Semantic Error: semantically similar to the correct parameter name. 
It is difficult for rule-based methods to locate the error types of value, and we come up with \textbf{E4.1}: Type Error: The parameter type is inconsistent with the API documentation. 

Figure \ref{error case} shows that the main source of API errors is actually E2.1, where LLMs misinterpret user intent and choose other candidate APIs. API requests generated by general LLMs (LLaMA-2-7B and GPT-3.5 Turbo) are difficult to parse correctly (E1). There are some formatting of the API name existing in LLaMA-2-7B, such as E2.2 and E2.3. The error of invalid parameter names (E3) are relatively few. The other errors in E4 take up a large portion, as there is no ground true for parameter value, making it difficult to accurately classify error types.

These predefined error types would be used as pseudo-facts for feedback, which greatly diminish the hallucinations of the LLM in API request generation. 
In the experiment section, this point will be illustrated by comparing the effects of different granularity error types on static feedback.

\section{METHODOLOGY}
\subsection{Overview}
Our Framework, AutoFeedback contains two low-coupled and complementary components: Static Scanning Component (SSC) and Dynamic Analysis Component (DAC). We combine it with the \textit{origin process (gray arrows)} of LLM calling API, creating two feedback loops, as illustrated in Figure \ref{fig:framework}. To avoid the dead loop, we set the maximum number of feedback as a hyperparameter.

\textit{1. Static feedback (green arrows).} The LLM takes the API documentation and user instruction as input, parses its generated output to get the API request. SSC perform a local static scanning to determine if any error exists in the API request. If none, proceed with sending the request; if any, conduct error detection to locate sub-errors, and give the precise static feedback to the LLM.

\textit{2. Dynamic Feedback (purple arrows).} After the API request is sent to the server and executed, the user assesses whether the returned API response meets his requirement. For those that failed, the DAC retrieves their detail error message in the API documentation and combines with the API response as dynamic feedback.

\textbf{Component Input:} The input of SSC includes user instruction, API documentation (including API name and its description, the type and description of its required parameter, exception specification), and the API request in the format \textit{APINAME(key1=value1, key2=value2,...)} generated by the LLM; The input of the DAC additionally includes the API response from the API server.

Our framework also provides support for human-machine collaboration, recording all information in feedback as the logs to assist users in solving tasks.


\subsection{Static Scanning Component}
The LLM suffers from serious hallucinations \cite{hallucination}\cite{riseagent}, which lead to difficulties in generating API requests with standardized format input and accurate parameter values, further constraining their ability to accurately leverage external tools. 
To implements lexical constraints on the output of LLMs, some work restrict the decoder \cite{picard}\cite{lqml} to modify the probability distribution of output which often result in significant performance losses.
Other work activate the self-reflection ability in LLMs through elaborate prompts \cite{selfcheck}\cite{deliberate}\cite{reflexion}. But these approaches only provide static feedback from the output of LLMs, which may include hallucinations and lack of the factual information. 

SSC categorizes errors into predefined types in Section 2, providing the factual information about the error location and error cause, significantly increasing the validity and accuracy of static feedback.

\subsubsection{API Retriever.} In Section 2, a major fraction of the cases in E2 are mistakenly selected other candidate APIs (E2.1). To deal with this problem, we use a deep learning model as an API retriever $M$, which could calculate the semantic similarity between two texts.
\begin{equation*}
M(text_1, text_2) = S, \text{      }, S \in [0, 1].
\end{equation*}
The model encodes the user instruction and the API description in documentation into two embeddings, and calculates their relevance with embedding similarity. Then, we ranks APIs in documentation based on the similarity score between the description and the instruction. Finally, SSC list the top $k$ scoring APIs (generally set $k$ to 1) as relevant APIs set $R$, which is subset of the API documentation. The API names in $R$ are consider to meet user needs, and will be provided for Error Dection sub-component.

\subsubsection{Error Detection.}
The LLM output may not contain API requests or cannot be parsed, which could be summarized as E1.
For other generated API requests, SSC perform automated error detection as follow:

\textbf{API name (E2.1, E2.2, E2.3).}
First SSC determines whether the API requests $r^1$ generated by the LLM is in $R$. If not ($r^1 \notin R$), the API name is incorrect (E2). And the next is to identify the error source. If the API name appears in the API documentation, the sub-error type is E2.1. This is considered to be a misunderstanding of user intent by the LLM, which resulted in selecting other APIs incorrectly. Because of the different in code styles, the LLM incorrectly generates the name of the underscore (e.g. \textit{user\_login}), whose real value should be the hump method (\textit{userLogin}).
SSC changes all API names to lowercase and remove characters other than alphanumeric by regular expression (\verb|r'[^a-zA-Z]'|). If the API $r^2$ in API documentation and $r^1$ have the same name after modification, the sub-error type is E2.2.
The detection for E2.3 is primarily aimed at LLMs making up non-existent and fake APIs in hallucinations. For example, the user instruction is \textit{"I'm trying to find out how much aspirin is left."}, and the generated fake API is \textit{"find\_aspirin\_number()"} while the real API is \textit{"list\_medicines(name='aspirin')"}. 
We also use $M$ to detect semantic similarity scores between the name of $r^1$ and the name of candidate APIs in API documentation. If there is an API $r^3$ that causes the score to exceed the threshold (generally set to 0.5), the sub-error type is E2.3. Other situations are the other error in E2.

\textbf{API parameter name (E3.1, E3.2, E3.3).}
When $r^1 \in R$, SSC next verifies the correctness of the API parameter name. Unlike the API name, we already know the correct parameter names, as they are provided in the API documentation. If the parameter name in $r^1$ does not exist in the corresponding description in the API documentation, which serves as E3. We consider this incorrect API parameter as $p^1$. If the name of $p^1$ appears in the parameter list of other APIs in the document, the sub-error type is E3.1. SSC perform the same operation on API parameter names as in E2.2 and E2.3 for API name. The sub-error type is E3.2 when $p^1$ matches the parameter $p^2$ in other API after modification. If the similarity score between parameter $p^3$ in API documentation and the name of $p^1$ is greater than the threshold, it is considered E3.3. Other situations are the other error in E3.

\textbf{API parameter value (E4.1).}
Due to the great flexibility in parameter values, we could only make preliminary judgments based on the parameter type.
SSC parses all parameter values in $r^1$ and compare them with the standardised data types in the API documentation. The sub-error type is E4.1 when the match of the parameter value $v^1$ is incorrect and the corresponding parameter serves as $p_v$. In some case, there are no errors in after static scanning, but existing during the calling process, or the calling results still cannot meet the user's requirements, it is the other error in E4.

\textbf{Return values.}
SSC performs error detection in ascending order of number (e.g. E1->E2, E2.2->E2.3,), and once an error is detected, return the corresponding value to enter static feedback. Besides for the error type, the other return values are displayed in Table \ref{tab:outputs}. 

\begin{table}[!t]
\renewcommand\arraystretch{1.0}
\setlength{\abovecaptionskip}{0cm}
\setlength{\belowcaptionskip}{-0.cm}

\vspace{-0.3cm}
\centering
  \caption{\small The return value of Error Detection.}
  \label{tab:outputs}
\footnotesize
\begin{tabular}{c | ccccc}
\toprule\toprule

Error Type    & Return                              \\ \midrule
E1            & -                                   \\ \midrule
E2.1 \& other & the name of $r^1$                   \\
E2.2          & the name of $r^1$,the name of $r^2$ \\
E2.3          & the name of $r^1$,the name of $r^3$ \\ \midrule
E3.1 \& other & the name of $p^1$                   \\
E3.2          & the name of $p^1$,the name of $p^2$ \\
E3.3          & the name of $p^1$,the name of $p^3$ \\ \midrule
E4.1 \& other & $v^1$, the description of $p_v$      \\
\bottomrule\bottomrule
\end{tabular}
\vspace{-0.3cm}
\end{table}

\begin{figure}[htbp]
\setlength{\abovecaptionskip}{0cm}
\setlength{\belowcaptionskip}{-0.cm}
  \centering
  \includegraphics[width=0.5\linewidth]{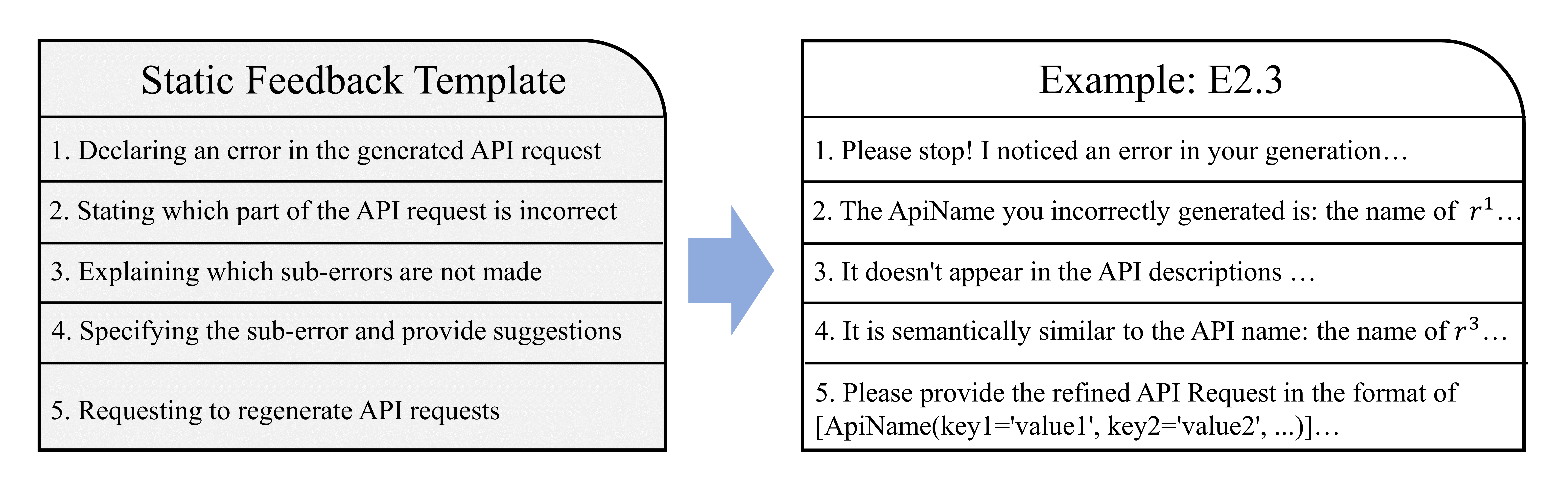}
  \caption{\small{Static feedback template and its example of E2.3.}}
  \label{fig:template}
  \vspace{-0.3cm}
\end{figure}

\subsubsection{Static Feedback.}
SSC receives the generated API request as input, and with the assistance of the API retriever, conducts error detection, resulting in static error.
Then, SSC selects the appropriate prompt template based on the error type and completes it based on other return values to generate the static feedback. 
As illustrated in Figure \ref{fig:template}), the template is roughly divided into five parts, with the granularity of feedback becoming finer sequentially.
Firstly, we declare that there is an error in the API request; Secondly, inform the error location and content; Thirdly, eliminate sources of errors that have already passed detection (for E2.3, E2.1 and E2.2 supposed to be excluded); Then, locate the source of the error and provide suggestions for correction; Finally, request the LLM to regenerate API requests. The static feedback of some errors might be missing certain parts. For instance, in E1, due to the first detection, there is no third part.

\subsection{Dynamic Analysis Component}
DAC focuses on the fact that many APIs return insufficient information (e.g. error code), after execution at the server, rather than detailed error messages (as shown in Figure \ref{figure-example}). In this case, the user have to continue querying the API documentation to find a detailed explanation of the corresponding error code, which is user-unfriendly and inconvenient. DAC automates the retrieval process, increases the detail level of dynamic feedback, and facilitates to correct corresponding errors in the API request.

\subsubsection{Document Retrieval.} DAC retrieves the details of the corresponding error from API documentation. We take the Retrieval Augmented Generation (RAG)\cite{rag} approach to extract the error-related message.

\begin{figure}[htbp]
\setlength{\abovecaptionskip}{0cm}
\setlength{\belowcaptionskip}{-0.cm}
  \centering
  \includegraphics[width=0.5\linewidth]{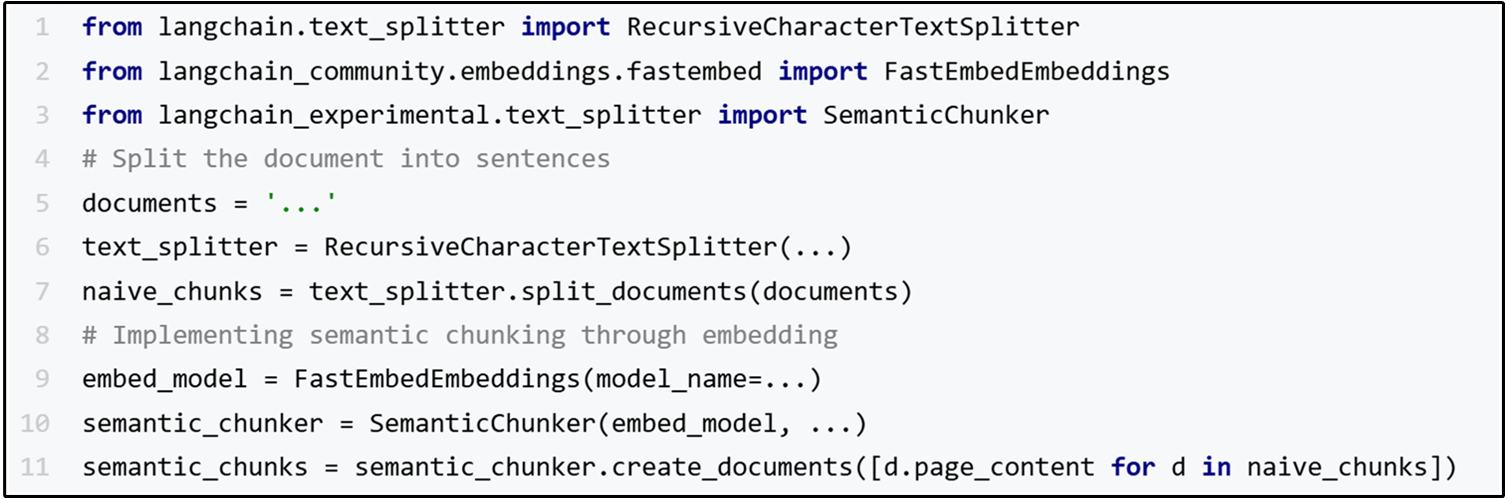}
  \caption{\small{Code example of documentation chunking.}}
  \label{fig:ragcode}
  \vspace{-0.3cm}
\end{figure}

The relevant code example is shown in the Figure \ref{fig:ragcode}. We first split the API documentation to obtain all document content related to the API (including functional descriptions, exception specification, etc.). Then we split these contents into sentence granularities, and use the embed model to vectorize sentences. Next, we places semantically similar sentences into the same data block, achieving semantic chunking and vectorization. This is equivalent to building a vector indexed text database, where vectorized sentences serve as indexes.

\begin{figure}[htbp]
  \vspace{-0.2cm}
\setlength{\abovecaptionskip}{0cm}
\setlength{\belowcaptionskip}{-0.cm}
  \centering
  \includegraphics[width=0.5\linewidth]{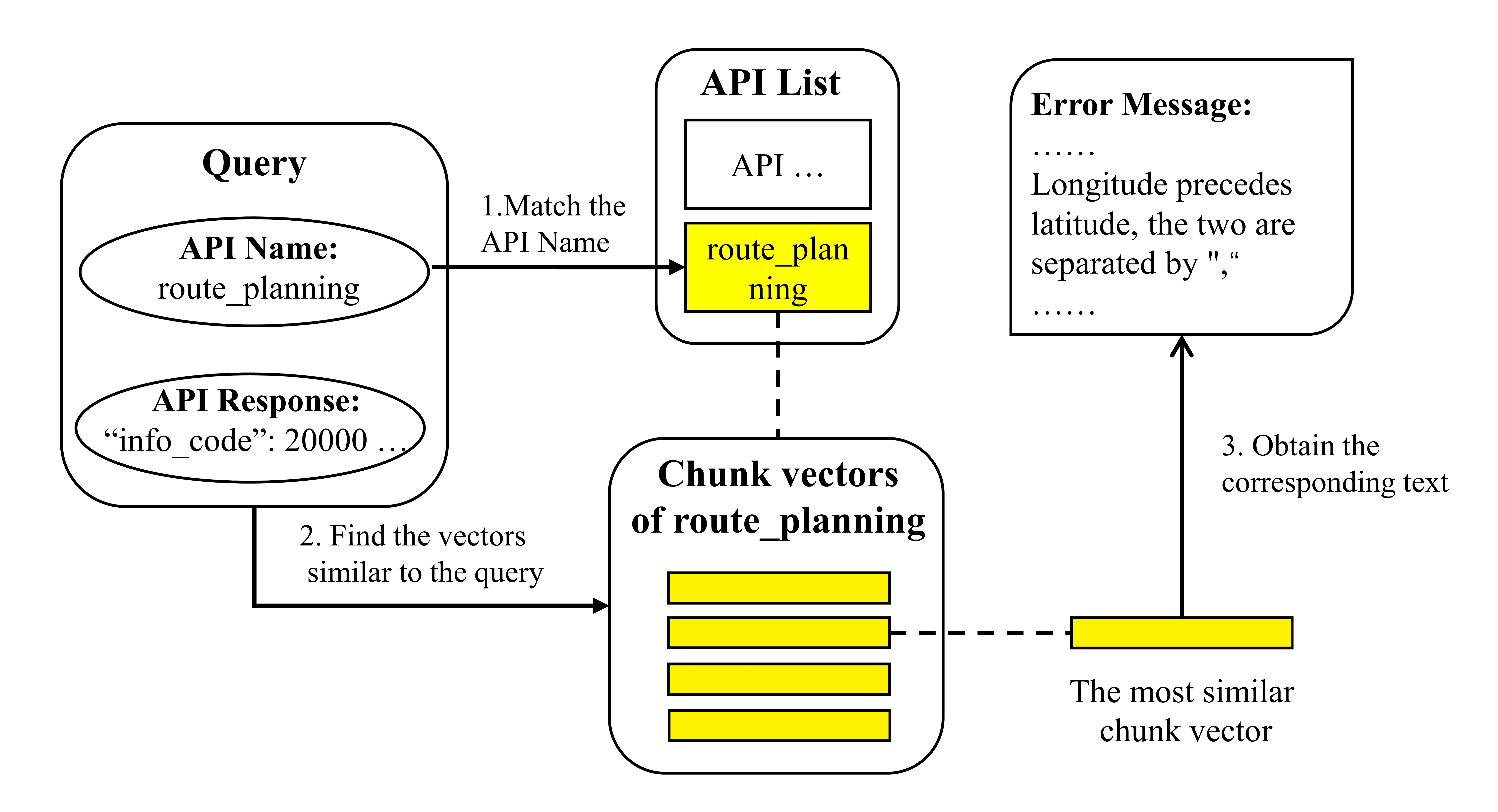}
  \caption{\small{The retrieval process of the error message.}}
  \label{fig:ragprocess}
  \vspace{-0.3cm}
\end{figure}

During the retrieval process in Figure \ref{fig:ragprocess}, DAC uses the API name (e.g. route\_planning) to match corresponding chunk vectors. Then use the same embed model to vectorize user query, and find for the most similar chunk vector as an index (implemented by approximate nearest neighbor algorithm), to retrieve the corresponding text from the database constructed earlier. Finally, a more detailed error message is obtained, such as the "Longitude precedes latitude", instead of "info\_code:20000" in Figure \ref{figure-example}.

\subsubsection{Dynamic Feedback}
The LLM uses its own internal representations to produce outputs which is not grounded in the external world, while it's crucial for the tool-augmented LLM to interaction with external tools, which could mitigate the possibility of the fact hallucinations.
Therefore, our framework integrates ReAct algorithm\cite{react} into the DAC, which allows LLMs to observe the API response to incorporate factual information into reasoning.
Simultaneously, we combined the error message from the Document Retrieval sub-component with the API response to form the dynamic feedback.

\begin{algorithm}[htbp]
\caption{Dynamic Feedback Process}
\label{algorithm:1}
\begin{algorithmic}[1]
\REQUIRE API Request $r$, API Documentation $doc$, maximum number of feedback $N$
\ENSURE API Response $P$
\STATE $i = 0$ 
\STATE $Record = []$ \COMMENT{Feedback Records}
\STATE $P \gets API\_Server(r)$ \COMMENT{API Execution}
\WHILE{$\textbf{not } Meet\_Requiremnet(P) \textbf{ } \OR \textbf{ } i \leq N$}
    \STATE $e \gets Document\_Retrieval(r, P, doc)$ \COMMENT{Error Message}
    \STATE $action \gets r$; $observation \gets (P, e)$
    \STATE $new\_action, thought = ReAct(action, observation, Record)$
    \STATE $Record_i \gets (action, observation, thought, new\_action)$
    \STATE $q \gets new\_action$
    \STATE $P \gets API\_Server(r)$
\ENDWHILE
\RETURN $P$
\end{algorithmic}
\end{algorithm}

The overall process of the Dynamic Feedback is shown in the algorithm \ref{algorithm:1}. The server executes the sent API request $r$ and obtrains the corresponding API response $P$. The next step is to assess whether the requirement of user have been met or if the maximum feedback count $N$ has been reached. Returns $P$ directly if the condition is satisfied. Otherwise, it enters the dynamic feedback loop: DAC first retrieves the corresponding error message in the $doc$; The API request $r$ serves as the action and API response and error message are combined as the observation in ReAct, which are fed into LLM together with $Record$; Newly generated API request and the thought will be stored with the action and the observation as a feedback record for the next feedback. Finally, we assess on the server's new execution result.

\subsection{Complementarity of SSC and DAC}
SSC is primarily oriented towards aiding in the generation of API requests that adhere to the user instruct and formatting standards. 
It's worth mentioning that SSC couldn't give feedback for details related to the parameter values. However, this does not imply that SSC is not beneficial for correcting parameter values, as it narrows down the sources of API errors, facilitating static and dynamic feedback to focus on parameter values. Our experiments demonstrated that the combination of SSC and DAC could significantly improve performance far beyond that of either one individually.

\begin{table}[]
\renewcommand\arraystretch{1.0}
\setlength{\abovecaptionskip}{0cm}
\setlength{\belowcaptionskip}{-0.cm}
\centering
  \caption{Comparison of Accuracy (\%) between no feedback (Base) and with feedback.}
    \label{tab:main results}
\resizebox{1.0\linewidth}{!}{
\begin{tabular}{ c |c c c | c | c c c c | c c c c }
\toprule\toprule
\multirow{2}{*}{\textbf{Method}}                                                               & \multicolumn{3}{c|}{\textbf{API-Bank \cite{apibank}}}               & \textbf{MP-API}\footnotemark[2]   & \multicolumn{4}{c|}{\textbf{ToolAlpaca-single \cite{toolalpaca}}}     & \multicolumn{4}{c}{\textbf{ToolAlpaca-mix \cite{toolalpaca}}}      \\   
& \footnotesize{LLaMA-2-7B}     & \footnotesize{Mistral-v0.2-7B} & \footnotesize{GPT-3.5 Turbo} & \footnotesize{GPT-3.5 Turbo} & \footnotesize{LLaMA-2-7B}     & \footnotesize{ToolAlpaca-7B}  & \footnotesize{GPT-3.5 Turbo}  & \footnotesize{GPT-4 Turbo} & \footnotesize{LLaMA-2-7B}     & \footnotesize{ToolAlpaca-7B}  & \footnotesize{GPT-3.5 Turbo}  & \footnotesize{GPT-4 Turbo} \\ \midrule 
Base                                                                 & 47.87          & 70.18           & 62.52             &     42.48              & 34.48          & 68.97          & 70.69         & 84.48    & 32.83          & 64.18          & 73.88     & 75.37    \\ \midrule 
Static Feedback                                                      & \textbf{66.59} & \textbf{79.69}  & \textbf{83.56}  &   \textbf{61.48}                & 64.66          & 77.59          & 87.07     & 90.52        & 47.76          & 74.63          & 85.07      &  87.31   \\ \midrule 
Dynamic Feedback                                                     & \textbf{-}     & \textbf{-}      & \textbf{-}        & \textbf{-}        & 57.76          & 79.31        & 90.52     & 95.69        & 47.76          & 75.37          & 87.31    & 88.06      \\ \midrule 
\begin{tabular}[c]{@{}c@{}}Static \&\\ Dynamic Feedback\end{tabular} & -              & -               & -                 & -                 & \textbf{75.00} & \textbf{87.93} & \textbf{97.41} & \textbf{100.00}   & \textbf{61.19} & \textbf{89.55} & \textbf{94.03} & \textbf{96.27} \\ 
 \bottomrule\bottomrule
\multicolumn{4}{l}{ \small{$\bullet$ Best performance on metric is shown in bold.}}\\
\end{tabular}}
\vspace{-0.3cm}
\end{table}

\section{EXPERIMENTS}
We evaluate AutoFeedback using four research questions: 
\par\textbf{RQ1}: Does the AutoFeedback improve the accuracy of API request generation and effectively address four typical types of errors?

\par\textbf{RQ2}: Does AutoFeedback reduce the interaction cost of generating accurate API request with LLM?

\par\textbf{RQ3}: What is the performance contribution of each component in the framework?

\par\textbf{RQ4}: What challenges AutoFeedback face in practical applications?

\subsection{Experimental Settings}
 \subsubsection{Scenarios and Datasets}
To verify the effectiveness of AutoFeedback, our experimental settings comprises two scenarios: \textit{single-API calling} (1,2,3) and \textit{multi-API planning} (4). 
\begin{itemize}
    \item[1)] API-Bank\cite{apibank}. It is a well-known synthetic 
 API dataset designed for tool-using LLMs.
    \item[2)]
    MP-API\footnotemark[2]. It is a code dataset built on a Python library (Materials Project\footnote{\url{https://github.com/materialsproject/api}}), measuring the ability of LLMs to call materials chemistry related tools, annotated by human experts.
    \item[3)] ToolAlpaca-single\cite{toolalpaca}. We conducted task decomposition for each sample in the real-world API dataset ToolAlpaca, to ensure a single API request could complete the sub-task.
    \item[4)] ToolAlpaca-mix\cite{toolalpaca}. Due to API inaccessibility or other reasons, the number of available APIs in the original dataset is reduced. We construct the ToolAlpaca-mix by mixing the remaining functioning APIs in its simulated dataset.
\end{itemize}

In the first scenario, the model addresses the user's problem by calling a single API request. However, in realistic application scenarios, LLMs generally are supposed to plan how to combine multiple API calls to solve problems.

The general information about four datasets in shown in Table \ref{tab:dataset}. The average number of minimum API requests required in ToolAlpaca-mix is 1.38, the rest are single API datasets. ToolAlpaca-mix has the highest average number of parameters per sample at 4.51, while the rest in descending order are MP-API for 3.80, API-Bank for 2.18, ToolAlpaca-single for 1.99.

    
 \subsubsection{Experimental Subject}
To validate the effectiveness and generality of AutoFeedback, we have selected four widely used and advanced LLMs. Considering different usage, the models for our experiments include open and closed source, non-fine-tuning and fine-tuning models.
 \begin{itemize}
    \item[1)] LLaMA-2-7B\cite{llama2}. 
        It is one of most popular LLMs in the open-source community.
    \item[2)] Mistral-v0.2-7B\cite{mistral}. 
        It is one of the best performance models in the open-source community at the same scale.
    \item[3)] ToolAlpaca-7B\cite{toolalpaca}. 
        It is the same architecture as LLama-2-7B, fine-tuned on the ToolAlpaca training dataset \cite{toolalpaca}.
    \item[4)] GPT-3.5 Turbo\cite{gpt3.5}. 
    Published by Open AI, it is broadly used closed-source LLM in commercial applications.
    \item[5)] GPT-4 Turbo\cite{gpt4}. 
    Published by Open AI, it is currently one of the most powerful and intelligent LLMs.
\end{itemize}

 \subsubsection{Evaluation Metrics}
  \begin{itemize}
    \item[1)] \textit{Accuracy}. 
         The accuracy is calculated by dividing the number of API request samples that meet user requirements by the total number of samples. Due to the lack of API servers for API Bank and MP-API, we consider the generated API request that is identical to the standard answers as meeting user requirements. For the ToolAlpaca-single and -mix, we followed the approach in \cite{toolalpaca} which passes API requests and responses to GPT-4 for evaluation.
    \item[2)] \textit{Process Correctness}.
        For the scenario of multiple API calls (ToolAlpaca-mix), the call unrelated to the task could lead to unnecessary cost. We use it to measure the optimality of API request sequences.
        Same as \cite{toolalpaca}, we employ GPT-4 to assess whether the API requests generated by the LLM are consistent with the standard answer.
    \item[3)] \textit{Overhead}. 
        Existing commercial LLMs all charge for the number of tokens generated, and \cite{lqml} uses the token number to denote the generated overhead. We further measure the interaction overhead using $\frac{\frac{1}{\mathcal{N}} \sum_{i=1}^{\mathcal{N}}{token_i}}{Accuracy}$, which indicate the average overhead required for 1\% improvement in accuracy.
\end{itemize}
We used the same evaluation prompt in the paper \cite{toolalpaca} for \textit{Accuracy} and \textit{Process Correctness}. In addition, we selected three experts to evaluate the results under the same conditions to demonstrate the effectiveness of GPT-4 evaluation.

\subsubsection{Experiment Environment}
We implemented AutoFeedback using PyTorch. We run all experiments on Ubuntu 20.04. The experiments were carried out on a machine with one Intel(R) Core(TM) i9-10850K CPU @ 3.60GHz, 32 GB main memory and one NVIDIA GeForce RTX 3090. The default API Retriever\footnote{https://huggingface.co/sentence-transformers/all-MiniLM-L6-v2} and the embed model\footnote{https://huggingface.co/BAAI/bge-small-en-v1.5} in AutoFeedback are lightweight sentence-transformers. For all experiments the maximum number of static feedback was set to 3, and the dynamic one was set to 2.


\subsection{Experimental Results and Analysis}
\textbf{RQ1: Does the AutoFeedback improve the accuracy of API request generation and effectively address four typical types of errors?} \\
We evaluated the most important metric - \textit{Accuracy} acorss a wide range of datasets and LLMs, illustrated in Table \ref{tab:main results}. The "Base" means the base performance without feedback. Since API-Bank and MP-API dataset do not have corresponding server to return responses, we have only made static feedback for them. 


The result demonstrates that AutoFeedback significantly improves the accuracy of various types of LLMs across all experimental datasets, reaching a maximum of 40.52\%. In particular, GPT-4 Turbo with AutoFeedback addressed nearly all problems in two real dataset scenarios, achieving 100\% on ToolAlpaca-single and 97.76\% on ToolAlpaca-mix;
GPT-3.5 Turbo, attained 83.56\% accuracy on API-Bank, 61.48\% on MP-API, and also solved almost all tasks on ToolAlpaca-single with 97.41\% accuracy and 94.03\% on ToolAlpaca-mix;
LLaMA-2-7B obtained 66.59\% accuracy on API-Bank compared to 47.87\% for the Base; for ToolAlpaca-single, it solved 75.00\% of the problems, far exceeding the Base of 34.48\%; similarly on ToolAlpaca-mix, it improved from 32.83\% to 61.19\%. In addition, Mistral-v0.2-7B increased its accuracy from 70.18\% to 79.69\% on API-Bank; as a fine-tuned LLM, ToolAlpaca-7B also achieved accuracy of 87.93\% and 89.55\% for ToolAlpaca-single and ToolAlpaca-mix, respectively, which is higher than the 68.97\% and 64.18\% at "Base".

To find out whether AutoFeedback accurately solves the typical errors, we used the API-Bank dataset as an instance to study the distribution of error types for GPT-3.5 Turbo. Meanwhile, to present the necessity of the feedback method, we first selected a non-feedback method as a comparison:
\begin{itemize}
    \item[1)] \textbf{LMQL}\cite{prompting} is the state-of-the-art language model programming method which improves the specification of content generated by LLMs via constraints.
\end{itemize}

The LMQL method only needs to predict the values in square brackets "[]": \textit{APINAME(key1=[value1], key2=[value2], ......)}. 

Then, to present the importance of detailed and factual feedback, we compared AutoFeedback with two representative and publicly available LLM coarse-grained feedback approaches:
 \begin{itemize}
     \item[2)] \textbf{DTG\cite{deliberate}} is a novel prompting framework. The LLM is self-reflective about its own generation through few demonstrations, detecting errors and correcting them.
     
    \item[3)] \textbf{Reflaxion}\cite{reflexion} is the state-of-the-art reinforce language agents framework. The external environment gives binary feedback to the LLM-based agent, i.e. success or failure.

    \item[4)] \textbf{AutoFeedback without Error Detection, (AutoFeedback w/o ED)}. Removing Error Detection sub-component in AutoFeedback, only gives feedback on relatively coarse-grained errors in API request, which are E1, E2, E3, E4.
\end{itemize}

\begin{table}[htbp]
\renewcommand\arraystretch{1.0}
\setlength{\abovecaptionskip}{0cm}
\setlength{\belowcaptionskip}{-0.cm}

\vspace{-0.2cm}
\centering
  \caption{\small Comparisons between different methods with GPT-3.5 Turbo on API-Bank.}
  \label{tab:baselines}
\resizebox{0.5\linewidth}{!}{
\begin{tabular}{c | ccccc}
\toprule\toprule
\textbf{Method}                                                      & \textbf{Accuracy (\%) $\uparrow$} & \textbf{E1(\%)$\downarrow$}     & \textbf{E2(\%)$\downarrow$}     & \textbf{E3(\%)$\downarrow$}     & \textbf{E4(\%)$\downarrow$}      \\ \midrule
\footnotesize{LMQL\cite{prompting}}                                                           & 54.35          & 13.99        & 2.23         & \textbf{0.00} & 29.37         \\ \midrule 
\footnotesize{DTG\cite{deliberate}}                                                            & 78.39          & 1.11         & 2.51         & \textbf{0.00} & 17.86         \\ \midrule 
\footnotesize{Reflaxion\cite{reflexion}}                                                      & 76.38          & \textbf{0.73} & 3.53         & 0.064        & 19.20         \\ \midrule 
\begin{tabular}[c]{@{}c@{}}\footnotesize{AutoFeedback}\\  \footnotesize{w/o ED}\end{tabular} & 81.02          & 0.99         & 1.59         & \textbf{0.00} & 16.27         \\ \midrule 
\footnotesize{AutoFeedback}                                                   & \textbf{83.57}  & 0.86         & \textbf{0.89} & \textbf{0.00} & \textbf{14.55} \\ \bottomrule\bottomrule
\multicolumn{6}{l}{ \small{$\bullet$ AutoFeedback w/o ED: AutoFeedback without Error Detection sub-component.}} \\
\end{tabular}}
\vspace{-0.2cm}
\end{table}

The result is shown in the Table \ref{tab:baselines}. Compared to all baselines, AutoFeedback achieved optimal performance (83.57\%). Apart from the slightly higher percentage of 0.86\% on E1 compared to the best Reflaxion's 0.73\%, its proportion across all categories of issues remained the lowest, effectively mitigating a spectrum of problems. Furthermore, with the granularity of feedback from coarse-to-fine, performance also gradually improves. For instance, Reflaxion gives binary feedback, AutoFeedback w/o ED gives coarse-grained feedback, and AutoFeedback gives fine-grained feedback. Relevant data for other models is the same tendency, and could be found in Appendix \ref{app: baselines}. 
\begin{figure}[htbp]
  \vspace{-0.2cm}
\setlength{\abovecaptionskip}{0cm}
\setlength{\belowcaptionskip}{-0.cm}
  \centering
  \includegraphics[width=0.5\linewidth]{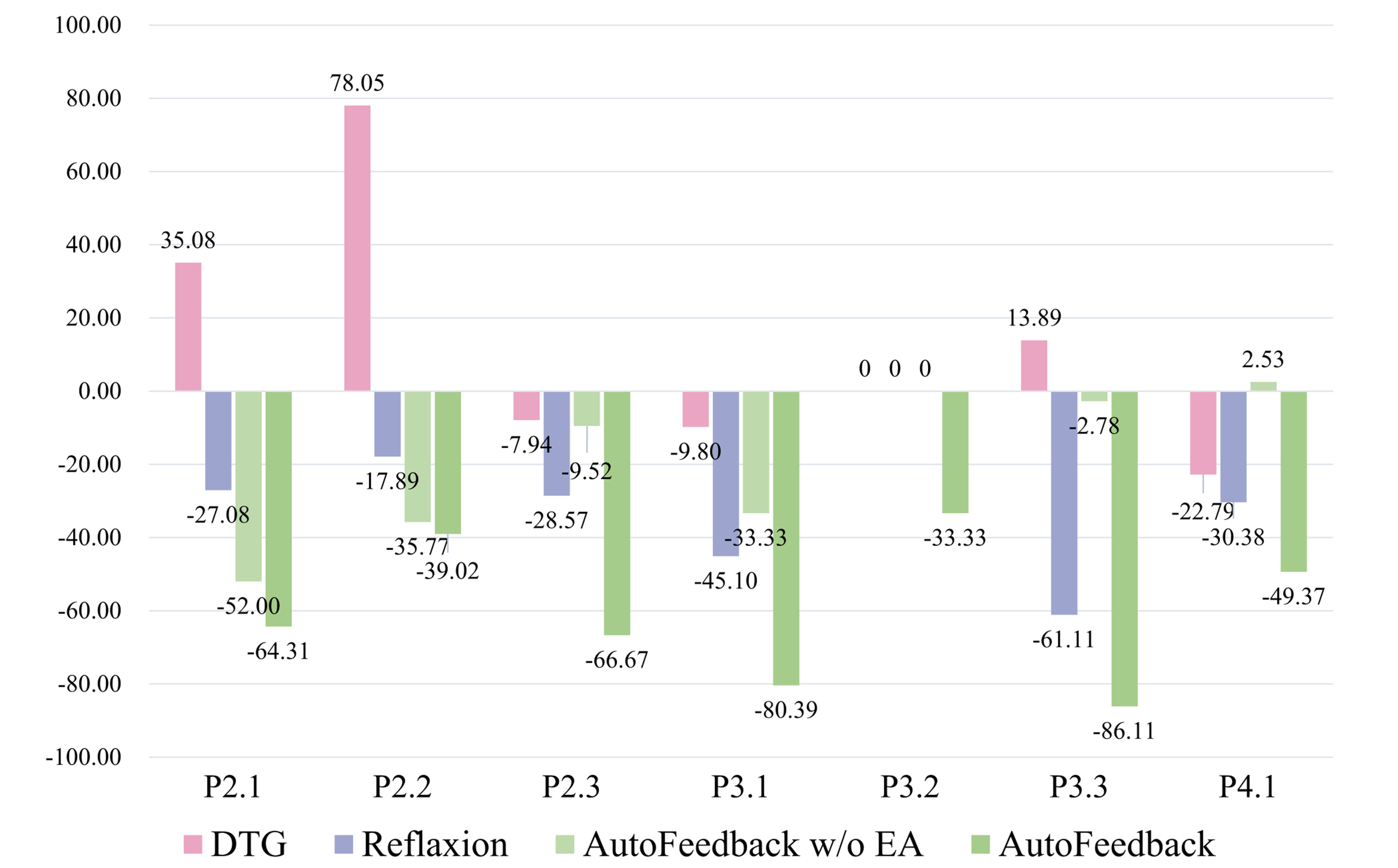}
  \caption{\small{The change rate (\%) of sub-errors after first iteration feedback in LLaMA-2-7B on API-Bank.}}
  \label{fig:detail}
    \vspace{-0.2cm}
\end{figure}

As a further step, we understand that the resolution of sub-errors at a more fine-grained level, using LLaMA-2-7B as an example, is shown in the Figure \ref{fig:detail}. AutoFeedback achieved the largest error decline rate on all types of errors which we could give accurate feedback. While DTG performed poorly and even added errors on E2.1, E2.2, E3.3. The reason for this discrepancy lies in DTG's lack of demonstrations in the initial feedback iteration, resulting in inaccurate self-detection error types. In contrast, the error types in AutoFeedback contain factual information. Similar to Table \ref{tab:baselines}, the finer the feedback granularity, the more the error rate decreases.

These experiments indicate that an increased provision of factual details through static feedback correlates with an enhanced precision in the generation of API requests by the LLM.

\begin{figure}[htbp]
  \vspace{-0.3cm}
    \centering
    \subfigure[Overhead of LLaMA-2-7B.
    \label{Llama-2-7B-cost}]
    {
        \centering
        \includegraphics[width=0.3\linewidth]{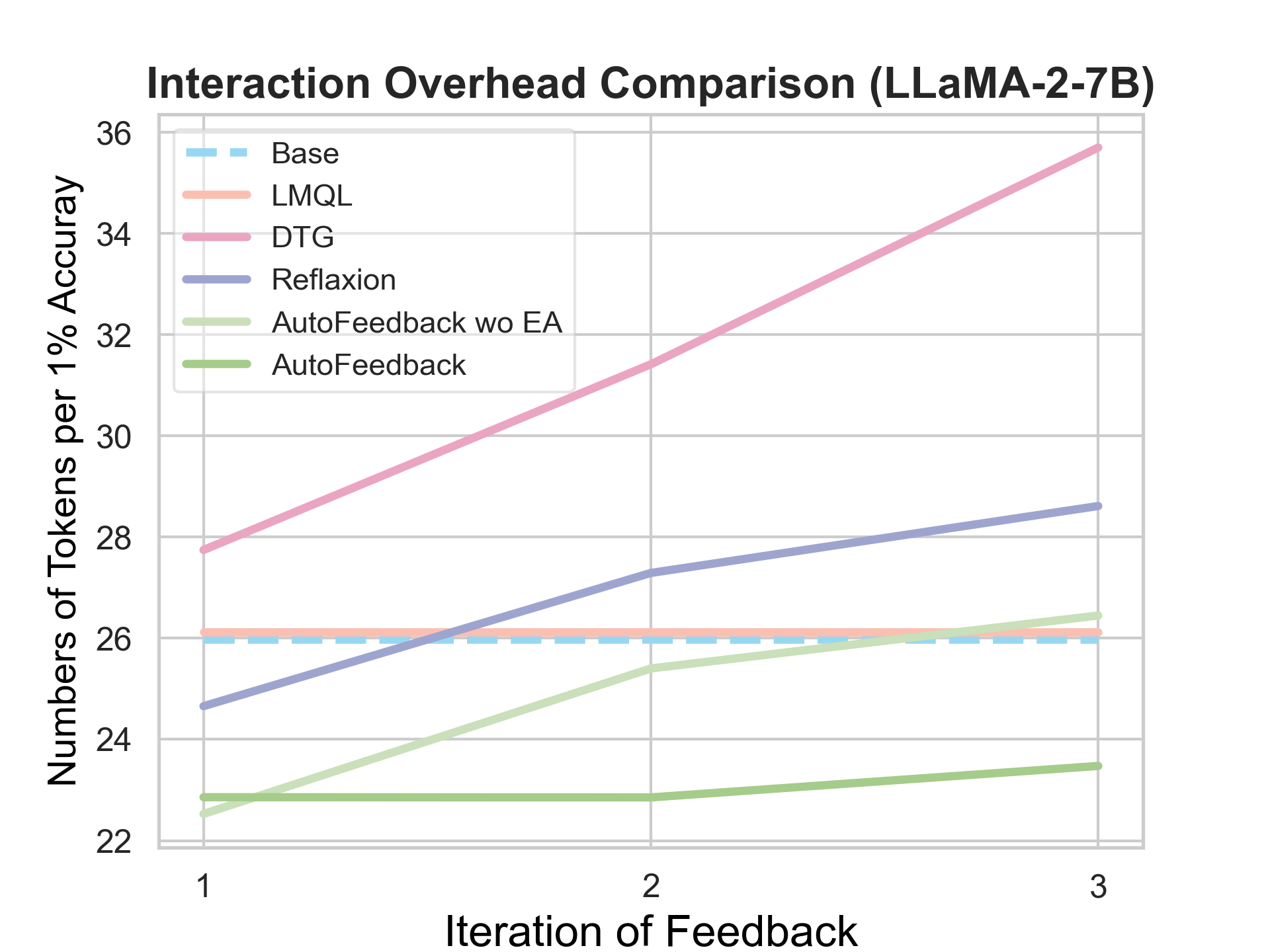}
    }
    \hfill
        \subfigure[Overhead of Mistral-v0.2-7B.
    \label{Mistral-v0.2-7B-cost}]
    {
        \centering
        \includegraphics[width=0.3\linewidth]{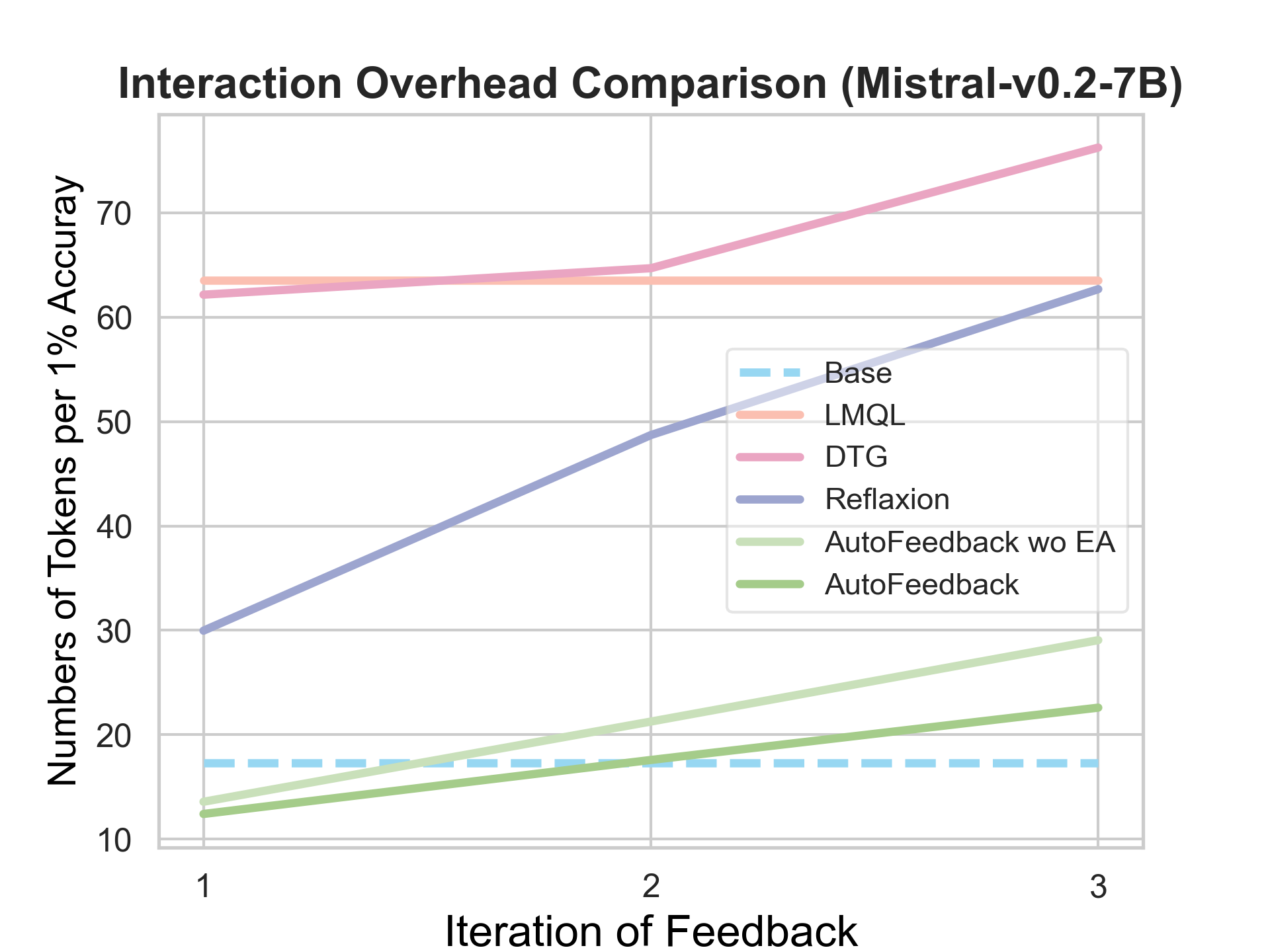}
    }
    \hfill
        \subfigure[Overhead of GPT-3.5 Turbo.
    \label{GPT-3.5 Turbo}]
    {
        \centering
        \includegraphics[width=0.3\linewidth]{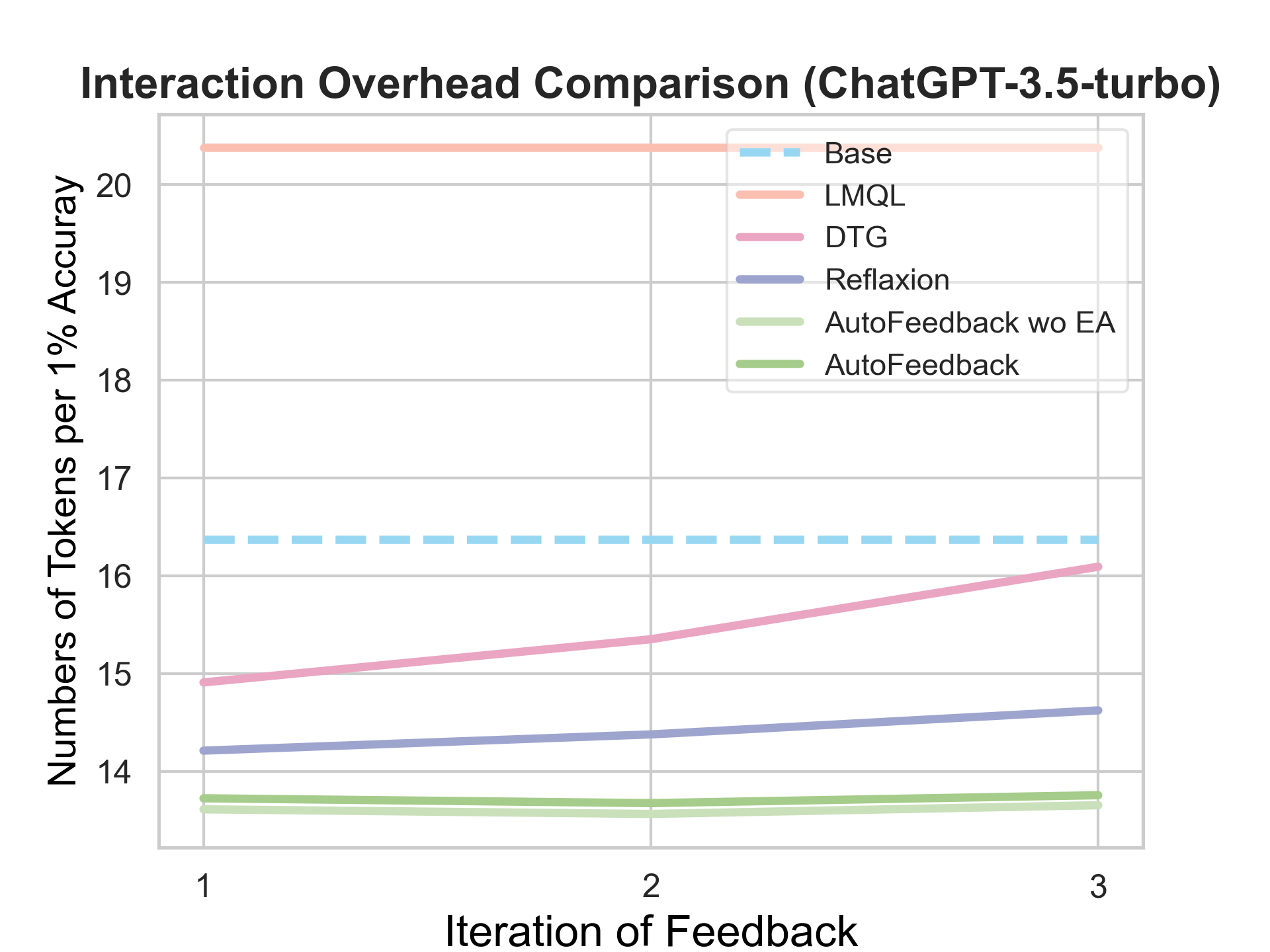}
    }
    \caption{Interaction overhead with different baselines on API-Bank}
    \label{fig:overhead}
      \vspace{-0.2cm}
\end{figure}

\textbf{RQ2: Does AutoFeedback reduce the interaction cost of generating accurate API request with LLM?} \\
We present the accuracy of ToolAlpaca-single and the interaction overhead in Table \ref{tab: overhead}. Although the average number of tokens for "Static \& Dynamic" feedback is the highest, due to its vastly elevated accuracy, the resulting interaction overhead is instead actually the smallest. For example, AutoFeedback has reduced the interaction overhead of GPT-3.5 Turbo from 13.01 to 9.96, a decrease of 23.44\%, GPT-4 Turbo from 10.97 to 9.67, resulting in 11.85\%.

The interaction overhead of three iterations of the LLMs on API-Bank is illustrated in Figure \ref{fig:overhead}. It could be observed that AutoFeedback maintains the lowest interaction cost under the same round of iterations on all LLMs. It is noteworthy that the interaction overhead tends to grow with the number of iterations increases. However, for highly intelligent LLMs such as GPT-3.5 Turbo, the overhead does not increase significantly.

\begin{table}[htbp]
\renewcommand\arraystretch{1.0}
\setlength{\abovecaptionskip}{0cm}
\setlength{\belowcaptionskip}{-0.cm}
\vspace{-0.3cm}
\centering
  \caption{\small Comparison of the feedback in Accuracy and Overhead on ToolAlpaca-single.}
  \label{tab: overhead}
\resizebox{0.5\linewidth}{!}{
\begin{tabular}{cc | ccc }
\toprule\toprule
\textbf{Model}                    & \textbf{Method}                                                                & \textbf{Avg. tokens} & \textbf{Accuracy (\%)$\uparrow$} & \textbf{Overhead $\downarrow$} \\ \midrule
\multirow{4}{*}{\footnotesize{\textbf{LLaMA-2-7B}}}        & \footnotesize{Base}                                                                    & 1054.47              & 34.48          & 30.58            \\
                                   & \footnotesize{Static}                                                                         &  1256.11		
          &   64.66         &    19.43      \\
                                   & \footnotesize{Dynamic}                                                                  & 1194.35              & 57.76           & 20.68          \\
                                   & \begin{tabular}[c]{@{}c@{}}\footnotesize{Static \& Dynamic}\end{tabular}     & 1338.03              & \textbf{75.00}  & \textbf{17.84}    \\ \midrule
\multirow{4}{*}{\footnotesize{\textbf{ToolAlpaca-7B}}}     & \footnotesize{Base}                                                                    & 1097.26              & 68.97           & 15.91             \\
                                   & \footnotesize{Static}                                                                         &  
       1139.28   &   77.59        &    14.68      \\
                                   & \footnotesize{Dynamic}                                                                  & 1178.75              & 79.31           & 14.86            \\
                                   & \begin{tabular}[c]{@{}c@{}}\footnotesize{Static \& Dynamic}\end{tabular} & 1196.10              & \textbf{87.93}  & \textbf{13.60}  \\ \midrule
\multirow{4}{*}{\footnotesize{\textbf{GPT-3.5 Turbo}}} & \footnotesize{Base}                                                                 & 919.45                  & 70.69           & 13.01           \\
                                   &\footnotesize{Static}                                                                           &     940.22		
           &    87.07        &     10.80      \\
                                   & \footnotesize{Dynamic}                                                                 & 990.84               & 90.52           & 10.94            \\ 
                                   & \begin{tabular}[c]{@{}c@{}}\footnotesize{Static \& Dynamic}\end{tabular} & 970.23               & \textbf{97.41}  & \textbf{9.96}     \\ \midrule
\multirow{4}{*}{\footnotesize{\textbf{GPT-4 Turbo}}} & \footnotesize{Base}                                                                 & 927.05                  & 84.48           &  10.97          \\
&\footnotesize{Static}                                                                           &     940.37		
&    90.52        &   10.39      \\
& \footnotesize{Dynamic}                                                                 & 960.04               & 95.69           &  10.03          \\
& \begin{tabular}[c]{@{}c@{}}\footnotesize{Static \& Dynamic}\end{tabular} & 967.11              & \textbf{100.00}  & \textbf{9.67}     \\
                                   \bottomrule\bottomrule
\end{tabular}}
\vspace{-0.2cm}
\end{table}

\textbf{RQ3: What is the performance contribution of each component in the framework?}

We first report the results of ablation study about the SSC and DAC components in Table \ref{tab:main results}.
Compared to the "Base", both "Static" and "Dynamic" feedback could improve accuracy, and their combination further enhance performance, far exceeding the effect of using them alone.

Next, regarding the Error Detection sub-components of SSC, we found that the absence of the sub-component in AutoFeedback ("AutoFeedback w/o ED") prevents it from accurately resolving errors in API requests, resulting in a performance degradation of 2.55\% in Table \ref{tab:baselines}. 

\begin{table}[htbp]
\renewcommand\arraystretch{1.0}
  \vspace{-0.2cm}
\setlength{\abovecaptionskip}{0cm}
\setlength{\belowcaptionskip}{-0.cm}
\centering
  \caption{\small Accuracy (\%) ablation experiments for sub-components in DAC with GPT-3.5 Turbo.}
  \label{tab: ablation dynamic}
  \resizebox{0.5\linewidth}{!}{
\begin{tabular}{c | cc}
\toprule\toprule
\textbf{Method}                                                     & \textbf{ToolAlpaca-single} & \textbf{ToolAlpaca-mix} \\ \midrule
Base                                                                & 70.69                      & 73.88                   \\ \midrule
\begin{tabular}[c]{@{}c@{}}Dynamic \\ w/o DR\end{tabular} & 84.48                      & 82.09                   \\ \midrule
\begin{tabular}[c]{@{}c@{}}Dynamic \\ w/o ReAct\end{tabular}        & 85.34                      & 80.60                   \\ \midrule
Dynamic                                                             & \textbf{90.52}             & \textbf{87.31}        \\ \bottomrule\bottomrule 
  \multicolumn{3}{l}{ \small{$\bullet$ Dynamic w/o DR: DAC without Document Retrieval sub-component.}} \\
\end{tabular}
}
\vspace{-0.3cm}
\end{table}

Ablation experiments on sub-components in DAC are shown in Table \ref{tab: ablation dynamic}. "Dynamic w/o DR" denotes providing dynamic feedback without the Error Message from Document Retrieval, and "Dynamic w/o ReAct" means that error messages and API responses are directly fed back to LLM. It could be observed that both "Dynamic w/o DR" and "Dynamic w/o ReAct" show worse performance compared to "Dynamic": the accuracy is reduced by 6.04\% and 5.18\% for ToolAlpaca-single; and is reduced by 5.22\% and 6.71\% for ToolAlpaca-mix. 
This indicates that Document Retrieval augments the volume of feedback information, whereas the ReAct algorithm activates the potential of LLMs to fully utilize feedback information. Relevant data for other models could be found in Appendix \ref{app: ablation}.

\begin{table}[htbp]
\renewcommand\arraystretch{1.0}
\vspace{-0.2cm}
\setlength{\abovecaptionskip}{0cm}
\setlength{\belowcaptionskip}{-0.cm}
\centering
  \caption{\small Comparison of the feedback in Process Correctness (\%) on ToolAlpaca-mix.}
  \label{tab: process}
  \resizebox{0.5\linewidth}{!}{
\begin{tabular}{c | cccc}
\toprule\toprule
\textbf{Method}           & \textbf{LLaMA-2-7B} & \textbf{ToolAlpaca-7B} & \textbf{GPT-3.5 Turbo} & \textbf{GPT-4 Turbo}\\ \midrule
Base             & 35.07               & 67.91                  & 79.10           &   85.07        \\ \midrule
Static           & \textbf{48.51}      & 76.12                  & 88.06          &   92.53         \\ \midrule 
Dynamic          & 38.06               & 70.15         & 87.31                 &  85.82   \\ \midrule
Static \& Dynamic & 47.76               & \textbf{77.61}         & \textbf{88.81}   &  \textbf{93.28}    \\ \bottomrule\bottomrule   
\end{tabular}
}
\vspace{-0.2cm}
\end{table}

We further investigate the impact of the role of static and dynamic feedback in the generation of API sequences, the result are presented in Table \ref{tab: process}.
We discovered that static feedback has the highest influence on process correctness improvement, while dynamic feedback has almost no effect. The combination of static and dynamic feedback improves problem solving accuracy, but it is not the optimal answer. As reflected in Table \ref{tab:main results}, the combination of static and dynamic feedback indeed improve accuracy in problem-solving. However, it appears that it doesn't enhance process correctness. 

This demonstrates that static feedback guides API requests towards the standard solution, selecting the optimal API, while dynamic feedback focuses on problem-solving, such as generating the correct parameter values. These two forms of feedback are mutually complementary in our framework.

\textbf{RQ4: What challenges AutoFeedback face in practical applications?}
 
\textit{1. API Documentation.} The most important prerequisite for AutoFeedback to operate successfully in real-world application is to have a standardised, descriptive and comprehensive API documentation. It is required that the documentation contains a correct description of the API functionality and its parameters. 
It is challenging for the fact that the API documentation is in a high-speed iterative updating process. For example, we have found in practice that some of the API tools in ToolAlpaca \cite{toolalpaca} have been lost due to a variety of issues, such as service inaccessibility, or API parameter change. The documentation is supposed to be updated in real-time as much as possible.

\textit{2. API Retriever.} Additionally, the selection of API retriever is also crucial, which is closely related to the resolution of the E2. In practical applications, we could utilize text information retrieval algorithm\cite{bm25} or fine-tune a pre-trained model, such as BERT\cite{bert}. 

\begin{figure}[htbp]
  \vspace{-0.2cm}
\setlength{\abovecaptionskip}{0cm}
\setlength{\belowcaptionskip}{-0.cm}
  \centering
  \includegraphics[width=0.5\linewidth]{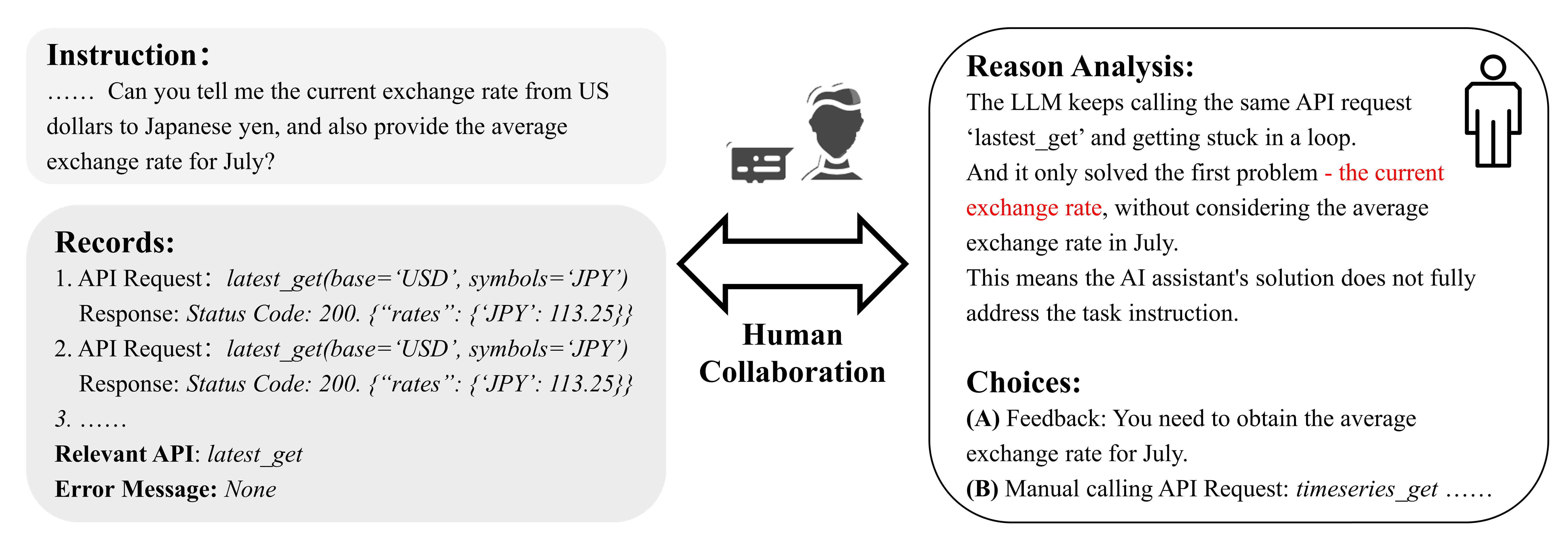}
  \caption{\small{
Human collaboration solving the feedback failure sample in ToolAlpaca-mix with GPT-4 Turbo.}}
  \label{fig:failure}
  \vspace{-0.3cm}
\end{figure}

\textit{3. Feedback Failure.} We found that with increasing feedback iterations, the accuracy is no longer increasing. Feedback does not improve the accuracy of API calls indefinitely, and the accuracy bottleneck remains closely related to model performance. When the set maximum number of feedback is reached but the problem remains unresolved, manual collaboration is required. In our experiments, GPT-4 Turbo cannot solve an sample in ToolAlpaca-mix, which is illustrated in Figure \ref{fig:failure}.

This sample involves Real-time and historical currency rates JSON APIs, which is a multiple API call scenario. GPT-4 Turbo trapped in a loop and repeatedly called the \textit{latest\_get} API to obtain the current Japanese yen exchange rate, without considering the average exchange rate in July. When using static feedback, we observed that the recommended relevant API was still \textit{latest\_get}, and no other static errors were identified. When looking at dynamic feedback, we found that each of its requests received a correct response (Status Code: 200), and therefore no error messages were present. Every time we give LLM feedback that the user's needs are not met, it cannot find the cause of the error, so it repeatedly calls \textit{latest\_get}. After analyzing the reasons, there are two different approaches could solve the problem: (A) Continue to provide feedback and request the LLM to get the average exchange rate for July; Or (B) manual call the corresponding API Request - \textit{timeseries\_get}. In summary, our framework provides information for users to collaborate when the feedback fails.

\subsection{Evaluation Effectiveness}


For the API-BANK and MP-API datasets, due to the lack of API servers, we compared them with the standard answers in the dataset to determine their correctness. This is a rigorous and effective evaluation method. However, API requests that are inconsistent with standard answers in real-world scenarios could still meet the user requirement. Therefore, for Toolalpaca-single and -mix datasets, we used the LLM as an agent to better evaluate the quality of API request generation. 

Although the LLM struggles to generate API requests, it is advisable to evaluate the results. As these are two tasks with different contexts, the evaluation task only requires the LLM to play the role of the user without requiring excessive professional knowledge. This method has been widely applied \cite{toolalpaca}\cite{eval1}\cite{eval2}\cite{eval3} and is considered to be more aligned with human preferences. We perform API parsing on the LLM output to avoid potential bias caused by direct evaluation by GPT-4.
Simultaneously, we invited three human experts to rate the experimental results to validate the effectiveness of the LLM evaluation.

\begin{table}[htbp]
\renewcommand\arraystretch{1.0}
\setlength{\abovecaptionskip}{0cm}
\setlength{\belowcaptionskip}{-0.cm}
\centering
  \caption{\small Comparison of the scores in Accuracy (\%) between human experts and the LLM.}
  \label{tab:expert}
\resizebox{0.5\linewidth}{!}{
\begin{tabular}{c|c|cc|cc}
\toprule\toprule
\multirow{2}{*}{\textbf{Model}}                           & \multirow{2}{*}{\textbf{Method}}                                     & \multicolumn{2}{c|}{\textbf{ToolAlpaca-single}} & \multicolumn{2}{c}{\textbf{ToolAlpaca-mix}} \\ 
                                                 &                                                             & $Human_{(var)}$      & LLM        & $Human_{(var)}$            & LLM            \\ \midrule
\multirow{2}{*}{GPT-3.5 Turbo}                    & Base                                                        & $72.41_{0.0230}$                       &    70.69          &  $73.88_{0.0166}$                 & 73.88                \\ 
                                                 & \begin{tabular}[c]{@{}c@{}}Static \& Dynamic\end{tabular} & $96.84_{0.0038}$                       &   97.41           &    $93.35_{0.0033}$              & 94.03                \\ \midrule
\multirow{2}{*}{GPT-4 Turbo} & Base                                                        &   $83.62_{0.0096}$                  &   84.48           &     $74.88_{0.0132}$            &   75.37             \\ 
                            & \begin{tabular}[c]{@{}c@{}}Static \& Dynamic\end{tabular} &    $98.85_{0.0077}$                  &  100.00           &      $96.77_{0.0033}$           & 96.27               \\
\bottomrule\bottomrule
\end{tabular}
}
\vspace{-0.2cm}
\end{table}

Human experts rate each sample on {0, 1}, where 1 represents the API requests meeting requirements. As presented in Table \ref{tab:expert}, we calculated the average scores and variances of three human experts on each sample and compared them with the scores of the LLM (GPT-4). It is evident that the scores assigned by human experts closely align with those generated by the LLM, with both exhibiting minimal variance (the variance is 0.2222 for the ratings [1, 0, 0] and [1, 1, 0]).

\begin{figure}[htbp]
  \vspace{-0.2cm}
\setlength{\abovecaptionskip}{0cm}
\setlength{\belowcaptionskip}{-0.cm}
  \centering
  \includegraphics[width=0.5\linewidth]{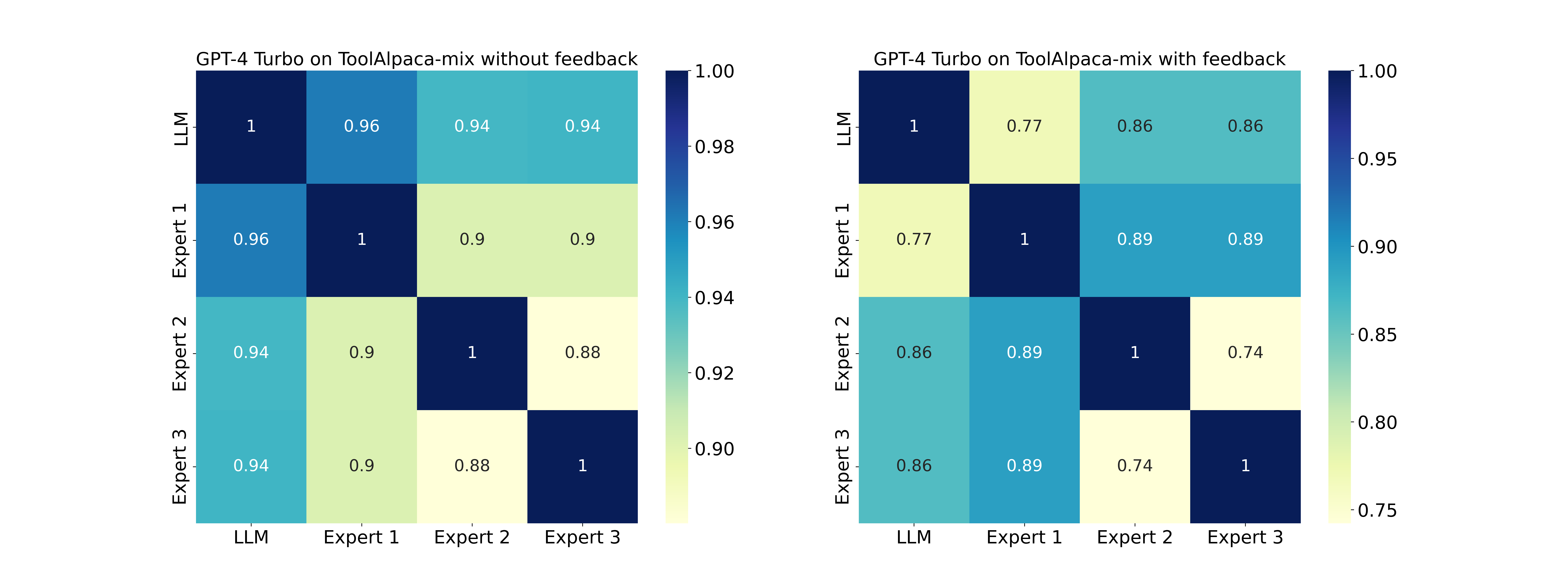}
  \caption{\small{
The Spearman Coefficient $\rho$ between the scores of human experts and the LLM.}}
  \label{fig:spearman}
  \vspace{-0.3cm}
\end{figure}

We measure the correlation of evaluations between human experts and the LLM using \textit{Spearman coefficient} $\rho \in [-1, 1]$ \cite{fieller1957tests}.
The larger the absolute value of $\rho$ is, the more correlated it is. As observed in Figure \ref{fig:spearman}, the result is indicative of a high positive correlation between GPT-4 and human expert ratings, with $\rho$ ranging from 0.77 to 0.96 on the ToolAlpaca-single and -mix datasets.


\section{THREATS TO VALIDITY}
We have identified the following two major threats to validity: 

(1) \textit{Unpredictable API format: } The process of generating API requests in LLM is inherently uncertain. Even with specified formats being set in prompts, LLMs may not produce the desired output. Our practical experience suggests that the probability of the API request generation with standardized format could be greatly improved by adding special indicators to the prompt. For example, prompt the LLM to generate API requests between \textit{<<API>>} and \textit{<</API>>}, which also makes it more convenient to parse. 

(2) \textit{Feedback loop interrupt: }
Due to ambiguity in user instructions and API documentation or limitations in the LLM itself, AutoFeedback struggles solving problems in some cases, and could trap in an infinite feedback loop. In response, we have set up a mechanism for human collaboration. When reaching the preset maximum number of feedback iterations, AutoFeedback will break the loop to wait for the user to manually intervene, and all the information from the previous feedbacks is recorded as the logs for the user.

\section{RELATED WORK}
\subsection{Tool-augment LLM}
The utilization of external tools to extend the capabilities of LLMs has emerged as a rapidly growing research area \cite{augmented}. Tools commonly incorporated include web-browser \cite{webbro}, code interpreters \cite{toolqa}\cite{agentbench} and API \cite{restgpt}\cite{toolllm}\cite{toolalpaca}. Among them, the API is easy-to-use because its functionality is encapsulated and there is no need to pay attention to implementation details. Currently, tool-augment LLMs could be divided into two distinct categories: fine-tuned and non-fine-tuned. The fine-tuned LLM \cite{gorilla}\cite{toolalpaca}\cite{toolformer}\cite{apibank}\cite{talm} is tightly coupled to the specific tools in training datasets. For non-fine-tuned LLMs \cite{restgpt}\cite{toolllm}\cite{chameleon}, API requests are generally invoked in the form of a dialogue via prompts. From the experimental results, AutoFeedback provides the efficient feedback for the both.

\subsection{Prompt Engineering}
Prompt Engineering \cite{shot}\cite{chainofthough} is a novel area that focuses on the development and optimisation of prompts, which enabling efficient convergence of LLMs and other tools. 
To alleviate the hallucinations within LLMs, there are two main directions in Prompt Engineering. The first is to add as much real information to the interaction process as possible, including obtaining responses from the external environment \cite{react} and retrieves the fact from the factual database \cite{rag}\cite{ragorgin}. The second is to reduce the generation of hallucinatory content, which could be implemented by self-reflection \cite{reflexion}, and self-checking \cite{selfcheck}, self-detecting \cite{deliberate}. Through error detection and the feedback, AutoFeedback integrates both types of approaches.

\section{CONCLUSION}
Although the LLM has demonstrated remarkable intelligence, it is rarely successful in scenarios where it is combined with external tools, since LLM suffering from serious hallucinations. To alleviate this problem, this paper presents AutoFeedback, an LLM-based framework for efficient and accurate API Request Generation, with two components: Static Scanning Component (SSC) and Dynamic Analysis Component (DAC). Extensive experiments demonstrate that AutoFeedback not only improves the accuracy of API requests, but also reduces the interaction overhead of LLMs, in various scenarios with different types of LLMs. The multiple iterations of static and dynamic feedback in AutoFeedback have stimulated LLM's innate problem-solving abilities. AutoFeedback also maintains a record of the feedback which could be used as the logs for human collaboration.
We believe AutoFeedback could be a starting point for the practical application of LLMs.

\bibliographystyle{unsrt}  

\bibliography{references}





\end{document}